\documentclass{article}

\pdfoutput=1

\usepackage[nonatbib, final]{nips_2017}

\usepackage{amsmath} 
\usepackage[english]{babel} 
\usepackage{hyperref} 
\usepackage{listings} 
\usepackage{booktabs}
\usepackage{caption}
\usepackage{graphicx}
\graphicspath{ {C:/Users/Rafal/Desktop/draft_nips/} }
\usepackage{color}
\usepackage{amsthm}
\newtheorem{definition}{Definition} 

\title{Happiness Pursuit: Personality Learning \\ in a Society of Agents}

\author{
  Rafa\l{} Muszy\'nski \\
  University College London\\
  \texttt{r.muszynski@cs.ucl.ac.uk} \\
 \And
  Jun Wang \\
  University College London\\
  \texttt{jun.wang@cs.ucl.ac.uk}
}

\date{}

\begin{document}

\maketitle 

\begin{abstract}

Modeling personality is a challenging problem with applications spanning computer games, virtual assistants, online shopping and education.
Many techniques have been tried, ranging from neural networks to computational cognitive architectures.
However, most approaches rely on examples with hand-crafted features and scenarios.
Here, we approach learning a personality by training agents using a Deep Q-Network (DQN) model on rewards based on psychoanalysis, against hand-coded AI in the game of Pong.
As a result, we obtain 4 agents, each with its own personality. Then, we define happiness of an agent, which
can be seen as a measure of alignment with agent's objective function, and study it when agents play both against hand-coded AI,
and against each other. We find that the agents that achieve higher happiness during testing against hand-coded AI, have lower happiness when competing against each other.
This suggests that higher happiness in testing is a sign of overfitting in learning to interact with hand-coded AI, and leads to worse
performance against agents with different personalities.

\end{abstract}

\section{Introduction}

Personality is defined in \cite{schultz2016theories} as the ``unique, relatively enduring internal and external aspects of a person's character that influence behavior in different situations.''
Theories of personality strive to explain and describe different types of personalities among people.
Freud \cite{freudSE, freudEGO} was the first to develop a modern theory of personality based mostly on clinical observations. In \cite{schultz2016theories}, Freudian structure of personality is described as composing of three elements: id, ego, and superego. 
Id is concentrated on following basic instincts. It is interested in instant gratification, knows no morality and is selfish.
On the opposite side, superego constitutes the moral aspect of personality. It wants to act according to parental and societal values and standards.
Ego is the rational element of the Freudian model. Its role is to resolve the conflicts between the demands of id and the moralizing superego by using defense mechanisms such as denial or repression.
Many alternative approaches describing personality have been proposed since (e.g., see \cite{schultz2016theories}). Moreover, computational models of personality have also been implemented (see section \ref{sec:relWork}). Most of them concentrate on intrapersonal aspects of personality and work with hand-crafted features, in relatively simple settings.
 
In the last few years, the advances in Deep Learning (DL) \cite{lecun2015deep, schmidhuber2015deep} and the development of Deep Q-Networks (DQN) \cite{mnih2015} have made it possible to work with increasingly more complex environments, e.g. Atari 2600 games \cite{bellemare2013arcade}, using only pixels as input.

In this work, we explore the applicability of DQN to computational psychology on the intrapersonal level. We concentrate on id and superego components of personality. The key idea is to employ DQN to learn a Freudian-inspired personality model from raw input in the game of Pong and use the resulting agents with different personalities to study their performance both in training and in a society of bots.
Firstly, to achieve agents with different personalities, we formulate and train a DQN-based model using Freudian-inspired reward functions (section \ref{sec:model}). We let one DQN learn from a reward signal strengthening selfish behavior - in Pong this can be reflected by scoring a point against an opponent - which is a typical characteristic of id \cite{schultz2016theories}.
Another DQN learns superego - based on rewards corresponding to morality, cooperation. In the game, this can be represented by trying to win, but without getting ahead of our opponent by more than one point.
Secondly, we define a ``happiness'' measure of a resulting id/superego agent based on its performance in relation to its objective function (section \ref{sec:model}).
Finally, when the trained id and superego agents play against each other, we find that the agents that are less happy after the training session - or less aligned with their intended reward objective - are more happy when they compete against each other (section \ref{sec:exp}).
This suggests that less overfitting during training, leads to a more successful, resilient behavior of an agent in a society.

\section{Related work} \label{sec:relWork}

To our knowledge, the first attempt to computationally describe the psychoanalytic theory of mind was \cite{nitta1999modeling} (cf. \cite{nitta2002computational}), and used basic probability. We did not find any extensions of the model to computer simulations.
A computational model of personality using some of the traits of the Big Five theory of personality (the Five Factor Model, FFM; cf. \cite{schultz2016theories}) and neural networks was presented in \cite{read2010neural}.
Other studies using neural networks to learn a personality include \cite{poznanski2005changing,quek2007testing}.
Another probabilistic approach was given by \cite{kshirsagar2002multilayer}, which used a Bayesian Belief Network with the FFM to build a multilayer personality model in a chat application.
The fact that you can replicate many of the previous experiments used in computational models of personality, e.g. \cite{read2010neural, quek2007testing}, within a CLARION cognitive architecture was demonstrated in \cite{sun2014model}.
As far as other cognitive architectures are concerned, 
\cite{karimi2012computational} used one trait of the FFM in the ACT-R architecture and demonstrate the behavior of their model, PIACT, in a soccer simulation environment.
The BDI architecture was used to simulate personality in \cite{padgham1997system}, and more recently with the FFM in \cite{ahrndt2015modelling}.
The biggest difference in our approach to modeling personality is that it only involves specifying new reward signals for the id or superego components of the personality model to achieve complex behavior in a challenging environment.

\begin{figure}[!tbp]
	\centering
	\includegraphics[width=0.5\textwidth]{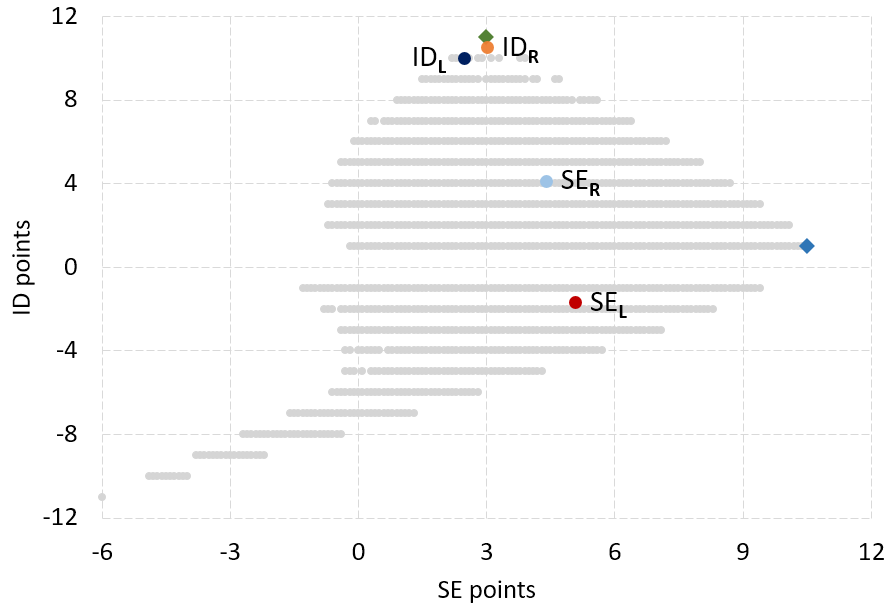}
	\caption{Possible $R_{ID}$ and $R_{SE}$ values combinations at the end of the game and the actual average values obtained by the agents during 1000 test games against hand-coded AI.}
	\label{fig:scenarios}
\end{figure}

\section{Problem description} \label{sec:model}

Motivated by the Freudian theory of personality \cite{freudSE, freudEGO}, we formulate two types of rewards and agents corresponding to the id and superego elements of personality.

\begin{definition}{ID reward ($r_{ID}$)}
is a predefined, scalar reward signal sent to the agent at each time step, encouraging selfish behavior.
\end{definition}

\begin{definition}{SE reward ($r_{SE}$)}
is a predefined, scalar reward signal sent to the agent at each time step, encouraging social behavior.
\end{definition}

\begin{definition}{Cumulative ID reward ($R_{ID}$)}
is a total reward achieved by an agent in a Markov Chain of length n, ending with a terminus event e:
\begin{equation}
R_{ID} = r_{ID1} + ... + r_{IDn} 
\end{equation}
\end{definition}

\begin{definition}{Cumulative SE reward ($R_{SE}$)}
is a total reward achieved by an agent in a Markov Chain of length n, ending with a terminus event e:
\begin{equation}
R_{SE} = r_{SE1} + ... + r_{SEn} 
\end{equation}
\end{definition}

\begin{definition}{ID}
is an agent with an objective of maximizing $R_{ID}$.
\end{definition}

\begin{definition}{SUPEREGO (SE)}
is an agent with an objective of maximizing $R_{SE}$.
\end{definition}

Happiness of people can be measured through surveys, e.g. \cite{happiness1999}, and captured as a scalar value on a scale. More recently, it has been described mathematically in computational neuroscience as a relation between certain rewards, expected values of given gambles, and the difference between expected and actual rewards in individual \cite{rutledge2014computational}, and more broadly in social context \cite{rutledge2016social}.
Our definition of happiness is for artificial agents and takes into account maximum, and minimum values they can obtain in an environment, and is independent of the happiness of other agents.

\begin{definition}{Happiness of Agent X ($H_X$)}
is defined as the quotient:
\begin{equation}
H_X = (R_X - R^*_X) / (R^{**}_X - R^*_X)
\end{equation}
where the cumulative reward obtained by agent X in a Markov Chain, ending with a terminus event e is denoted as $R_X$, and its potential maximum and minimum values as $R^{**}_X$ and $R^*_X$, respectively.
\end{definition}

Using the above definitions, we train ID and SE agents in the game of Pong.
We study their happiness both when training against hand-coded AI, and when they play against each other in a society of bots.

\begin{figure}[!tbp]
	\begin{minipage}[b]{0.485\textwidth}
	\centering
	\includegraphics[width=0.9\textwidth]{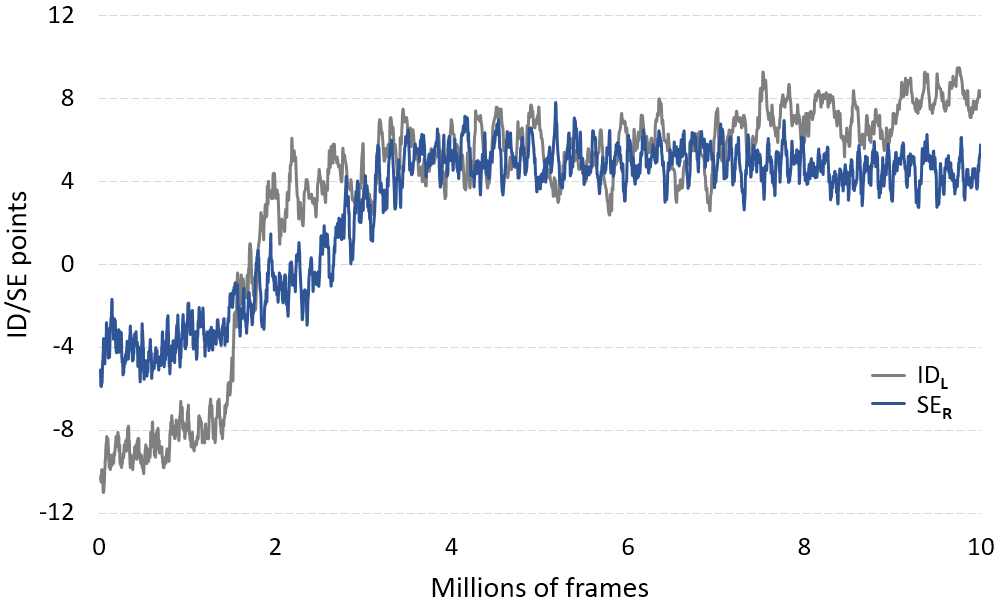}
	\caption{Training $ID_L$ and $SE_R$ agents against hand-coded AI for 10 million frames.}
	\label{fig:id_se_train}
	\end{minipage}
	\hfill
	\begin{minipage}[b]{0.485\textwidth}
	\centering
	\includegraphics[width=0.68\textwidth]{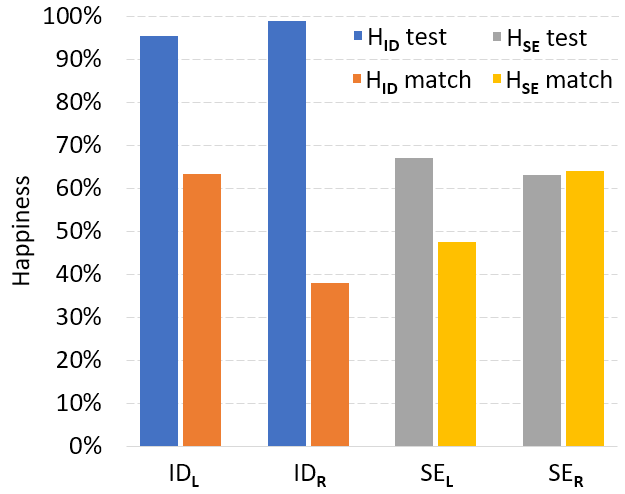}
	\caption{Lower happiness score during testing leads to higher happiness in a society.}
	\label{fig:happiness}
	\end{minipage}
\end{figure}

\section{Experiment set-up and results} \label{sec:exp}

In order to demonstrate $ID$ and $SE$ agents in practice, we modified a simple Pong game. The game has 2 players, each controlling one of the paddles on either side of the screen. The goal of the game is to bounce the ball in such a way, so that it goes past the opponent's paddle - for which the player is awarded 1 point. The game ends with one of the players scoring 11 points.

In terms of the terminology from section \ref{sec:model}, scoring 11 points is the terminus event ($e$), after which a new match begins.
A subjective reward ($r_{ID}$) used to train $ID$ relates to the id component of the Freudian theory of personality that is selfish. Here, we set $r_{ID}$ to +1, if the ball goes past the opponent ($ID$ scores a point), -1 if it goes past $ID$ ($ID$ loses a point), and 0 otherwise. As a result, the minimum value of $R_{ID}$ (i.e. $R^*_{ID}$) in one match is -11 (losing every single point), and the maximum is +11 ($R^{**}_{ID}$). On the other hand, $r_{SE}$ captures the social aspect of the environment, in line with the superego component of personality. Here, the rules for getting the reward are more involved. In short, the goal of $SE$ is still winning, ideally 11:10, and taking turns in scoring the points with the opponent.\footnote{To see the full code and the videos showing the performance of agents, go to https://git.io/vbT1v.} Hence, $R^*_{SE}$ is -6, and $R^{**}_{SE}$ is 10.5. Fig. \ref{fig:scenarios} shows possible values of $R_{ID}$ and $R_{SE}$ that an agent can obtain at the end of each match (for clarity, the values of $R_{SE}$ are rounded). Note the natural trade-off line between the optimal ID (pair $R_{SE}=3$, $R_{ID}=11$, the green diamond at the top)  and SE values (pair $R_{SE}=10.5$, $R_{ID}=1$, the rightmost blue diamond).

To train $ID$ and $SE$ agents, we use DQN \cite{mnih2015} with $r_{ID}$ and $r_{SE}$ as the reward signals. 
Both $ID$ and $SE$ agents are first trained separately against a hand-coded AI, and strive to maximize $R_{ID}$ and $R_{SE}$ respectively. We train $ID$ to control the paddle on the left and right on 10 million frames. We do the same for the $SE$ agent.
As a result we obtain 4 models: $ID$ trained to control the paddle on the left ($ID_L$), right ($ID_R$), and the $SE$ agents respectively ($SE_L$ and $SE_R$).
Note that $ID_L$ did not perform well when used to control the paddle on the opposite end of the screen (the same was observed for other agents), hence the 2 agents per side.
Fig. \ref{fig:id_se_train} shows a learning curve for the $ID_L$ and $SE_R$ agents during training. For clarity, the lines on the graph show the moving averages of $R_{ID}$ and $R_{SE}$ over the last 10 matches.
The learning curves of $ID_R$ and $SE_L$ look similarly and are therefore omitted.
Note that performance of $SE_R$ starts to deteriorate slightly around 4 million frames. We observe a similar behavior when training $SE_L$.
Hence, in the following experiments we froze the weights calculated at 4.2M frames for $SE_L$, 4.15M for $SE_R$, and 10M for both $ID$ agents.

In the test phase, we set the epsilon value to 0, and run each model for 1000 matches against the hand-coded AI (see Table \ref{table:test}).
One can note, that $ID$ agents learned to win every single time (\% won column in Table \ref{table:test}), with $ID_R$ agent beating the hand-coded opponent by a larger number of points than $ID_L$, as can be seen by its higher average $R_{ID}$ value.
$SE_L$ agent did not learn to win consistently, but it achieves the highest score among all the agents on its core objective - maximizing the $R_{SE}$ value.
The numbers indicate that $r_{SE}$ is a harder signal to learn, as both $SE$ agents did not get as close to their potential maximum ($R_{SE} = 10.5$, $R_{ID} = 1$) as both $ID$ agents.
The visual confirmation of this is presented in Fig. \ref{fig:scenarios}, where the average values of $R_{SE}$ and $R_{ID}$ obtained by the agents during testing are plotted against all the possible scenarios.
Indeed, the averages of both $ID$ agents are closer to their respective maximum (the green diamond at the top), than those for the $SE$ agents (the rightmost blue diamond).
We find that $r_{ID}$ is easier to learn as the same actions lead to the same rewards. The $SE$ agent has a more difficult function to learn, as the same actions result in different reward depending on the score. $ID$ rewards are independent of the score.

Subsequently, the 4 trained agents played 100 matches against each other (see Table \ref{table:matches}). The $ID_L$ agent - which performed worse than $ID_R$ during testing - won 91\% of the games against its opponent,
and also beat $SE_R$ by a small margin. $SE_R$ also achieved lower $R_{SE}$ than $SE_L$ in testing, yet it greatly outperformed $SE_L$ in a series of matches.
Finally, happiness statistics (see Figure \ref{fig:happiness}) further indicate, that agents achieving lower happiness on their respective objective during testing, show higher happiness when interacting with other agents during the matches.
This suggests that achieving high performance on the given reward objective against the hand-coded Pong AI is not enough to generate a resilient strategy. Agents may need more variability during training to better respond to the changing environment later on.

\begin{table}[]
\begin{minipage}[]{0.47\textwidth}
\caption{Performance of agents during 1000 test games against hand-coded AI.}
\label{table:test}
\begin{tabular}{lllll}
\toprule
                &   Average      &               & \multicolumn{2}{c}{Average} \\
Agent        &   score       & \% won        & $R_{ID}$            & $R_{SE}$ \\
\midrule
$ID_L$   &    11.00    & 100\%  & 9.99      & 2.48      \\
$ID_R$   &    11.00    & 100\%  & 10.76     & 3.03      \\
$SE_L$   &     8.84    & 16\%   & -1.69     & 5.09      \\
$SE_R$   &     10.78   & 91\%   & 4.08      & 4.41     \\
\bottomrule
\end{tabular}
\end{minipage}
\hfill
\begin{minipage}[]{0.5\textwidth}	
\centering
\caption{Summary statistics of 100 matches between agents.}
\label{table:matches}
\begin{tabular}{lllll}
\toprule
                  & \multicolumn{2}{c}{Average score} & \multicolumn{2}{c}{\% won} \\
Match         & L             & R            & L            & R \\
\midrule
$ID_L$ vs. $ID_R$ & 10.80        & 5.29       & 91\%         & 9\% \\
$SE_L$ vs. $SE_R$ & 3.50        & 10.93       & 3\%          & 97\% \\
$ID_L$ vs. $SE_R$ & 9.44        & 8.96       & 51\%         & 49\% \\
$SE_L$ vs. $ID_R$ & 8.92        & 9.05       & 40\%         & 60\% \\
\bottomrule
\end{tabular}
\end{minipage}
\end{table}

\section{Conclusions}

In this paper we describe the computational model of personality based on Freudian psychoanalysis and DQN, and define ``happiness'' of an agent.
Through the experiments, we show that the agents that are less aligned with their intended objective after the training period, exhibit more alignment when interacting with other agents. 
Here, we acknowledge the weaknesses of this work and mention possible further improvements.
Firstly, due to the nature of the DQN algorithm, it is impossible to compare our work with other computational models of personality as was done in \cite{sun2014model}.
Secondly, this study relies on the psychoanalytic theory of personality. However, it appears that extending this work to the FFM could further improve the ability of the model to capture human personality.
Lastly, in order to calculate the happiness, one needs to know the minimum and maximum values of $R_{SE}$ and $R_{ID}$, which may be impossible in more complex environments.

\section*{Acknowledgment}

This work was supported by Microsoft Research through its PhD Scholarship Programme.

\bibliography{untitled-2}

\end{document}